# Too Human to Model: The Uncanny Valley of LLMs in Social Simulation

*When Generative Language Agents Misalign with Modelling Principles*

Yongchao Zeng[1,*], Calum Brown [1,2], Mark Rounsevell[1,3,4]

[1] Institute of Meteorology and Climate Research, Atmospheric Environmental Research (IMK-IFU), Karlsruhe Institute of Technology, 82467 Garmisch-Partenkirchen, Germany

[2] Highlands Rewilding Limited, The Old School House, Bunloit, Drumnadrochit IV63 6XG, UK

[3] Institute of Geography and Geo-ecology, Karlsruhe Institute of Technology, 76131 Karlsruhe, Germany

[4] School of Geosciences, University of Edinburgh, Drummond Street, Edinburgh EH8 9XP, UK

* Corresponding author

E-mail address: yongchao.zeng@kit.edu (Y. Zeng)

**Abstract**：Large language models (LLMs) have been increasingly used to build agents in social simulation because of their impressive abilities to generate fluent, contextually coherent dialogues. Such abilities can enhance the realism of models. However, the pursuit of realism is not necessarily compatible with the epistemic foundation of modelling. We argue that LLM agents, in many regards, are too "human" to model: they are too expressive, detailed and intractable to be consistent with the abstraction, simplification, and interpretability typically demanded by modelling. Through a model-building thought experiment that converts the Bass diffusion model to an LLM-based variant, we uncover five core dilemmas: a temporal resolution mismatch between natural conversation and abstract time steps; the need for intervention in conversations while avoiding undermining spontaneous agent outputs; the temptation to introduce rule-like instructions in prompts while maintaining conversational naturalness; the tension between role consistency and role evolution across time; and the challenge of understanding emergence, where system-level patterns become obscured by verbose micro textual outputs. These dilemmas steer the LLM agents towards an "uncanny valley": not abstract enough to clarify underlying social mechanisms, while not natural enough to represent realistic human behaviour. This exposes an important paradox: the realism of LLM agents can obscure, rather than clarify, social dynamics when misapplied. We tease out the conditions in which LLM agents are ideally suited: where system-level emergence is not the focus, linguistic nuances and meaning are central, interactions unfold in natural time, and stable role identity is more important than long-term behavioural evolution. We call for repositioning LLM agents in the ecosystem of social simulation for future applications.

**Keywords:** Large language models, LLMs, Social simulation, Agent, Modelling.

# 1. Introduction

Large language models (LLMs), such as GPT-4 (Achiam et al., 2023) and Claude (2025), have been increasingly used to simulate human systems (Wang et al., 2024a; Zhang et al., 2024a). With the capability of generating coherent, human-like text, LLMs enable researchers to build and explore artificial societies comprising a variety of autonomous agents (Wang et al., 2024a). LLM-based agents can express intent, respond to requests, and even mimic complex social behaviours, which are often implemented through well-designed prompts and integration of multiple agents (Li et al., 2024b). This fosters the emergence of narrative-rich simulation (Aoki et al., 2023; Wang et al., 2024b; Yan and Xiang, 2025) – a system that can operate through natural language, rather than relying on predefined state machines or rule-based logic.

The development of LLMs is exciting. LLMs seemingly provide new features that conventional, rule-based agents often struggle with: plausible, situated, contextually rich behaviour. Literature shows that, through LLMs, simulation can capture the texture of human life (Aoki et al., 2023), ranging from daily conversations and social norms (Horiguchi et al., 2024) to improvisation and persuasion (Carrasco-Farre, 2024). Numerous recent papers demonstrate that in LLM-driven social environments, agents can negotiate dinner plans, discuss local events, or co-compose storylines (see Wang et al. (2024a) for a comprehensive review).

However, LLMs have many widely known technical limitations, such as hallucination (Li et al., 2024a; Perković et al., 2024; Tonmoy et al., 2024), bias (Gallegos et al., 2024; Lin et al., 2024; Tao et al., 2024), instability in formatting outputs (Liu et al., 2024a; Liu et al., 2024b; Zeng et al., 2025) – that may affect their integration with social simulation. Beyond these, we argue that there is a more profound methodological and epistemological tension underlying LLM-driven simulation, even if all these technical limitations are solved: LLM agents can speak and act like real humans, but this realism comes in social simulation by risking model validity, interpretability, and explanatory rigour. The core idea of this paper is that "LLM agents are often too human to model." Their simulation of human behaviours is too detailed to serve as an abstract representation, but too artificial to serve as an explanation. Their use, therefore, conflicts with the very nature of social modelling as an epistemic tool for extracting insights from the noisy real world (Edmonds, 2017; Epstein, 2012; Knuuttila, 2021; Miller and Page, 2009; Morgan, 2012; North and Macal, 2007).

This tension results from a fundamental mismatch between how models abstract social processes and how LLMs generate expressive behaviour. Where conventional agents evolve following clearly defined rules (Cuskley et al., 2018), LLM agents generate narratives conditioned on textual inputs (Aoki et al., 2023); where conventional simulation aims to distil mechanisms (Epstein and Axtell, 1996), LLM agents necessitate sophisticated external mechanisms designed to enable emotional change, memory management, and reasoning processes (Xie et al., 2024). Furthermore, while rule-based agents operate on abstract temporal frameworks (e.g., interacting yearly (Zhang et al., 2022)), LLM agents respond through high-resolution, turn-by-turn dialogues (Wang et al., 2024a). Such discrepancies make the causal linkages between microscopic agent behaviours and system-level social emergence unclear.

The purpose of this paper is to reflect critically on the trend of embedding LLM agents in social simulation without sufficient consideration of the methodological legitimacy. We use a classic innovation diffusion model – the Bass model (Bass, 1969) – to conduct a thought experiment demonstrating the consequences of the epistemic mismatch. Through the process of building an LLM-driven version of the Bass model, we elaborate on how LLM agents' realism gradually introduces liabilities, and how reducing such liabilities may result in an "uncanny valley of agents" – not abstract enough to clarify underlying social mechanisms while not realistic enough to naturally exhibit human behaviours.

We do not advocate abandoning LLMs in social simulation but to propose applying generative AI agents to domains where they fit better, such as situated role play (Zeng et al., 2024a), narrative framing (Wang et al., 2024b), and human-AI interaction (Tamoyan et al., 2024), rather than microscopic components intended to replace conventional agents. We call for the reframing of the role that LLMs can legitimately play in simulation science – one that respects the strengths of both abstraction and expression without confusing the two.

**2. A Brief Revisit of The Classic Bass Diffusion Model and Its Epistemic Strength**

Since Frank Bass first introduced his diffusion model in 1969 (Bass, 1969), it has been one of the most enduring, widely used approaches to modelling innovation diffusion processes. From predicting the popularisation of colour television (Bass, 1969) to modelling the adoption of renewable energy technology (Rao and Kishore, 2010), the Bass diffusion model provides a simple but powerful tool for researching how innovation, such as new thoughts, behaviours, technologies, and products, spread in society (Rogers et al., 2014). The model's power does not lie in its sophistication but in its simplicity and elegance (Bass, 2004): using only two core parameters to capture various diffusion dynamics in the real world. This model is certainly not flawless, but its structural clarity and interpretability make it a good illustrative example of how to build models. Before proceeding to build an LLM-based diffusion model, we briefly revisit the classic model here.

In the model's differential equation form, the proportion of new adopters $f(t)$ is calculated as:

$$f(t) = [p + q \cdot F(t)] \cdot [1 - F(t)],$$

where $F(t)$ is the cumulative proportion of adopters up to time t; p and q respectively represent:

1) External influence that causes individuals to adopt an innovation, such as the influence of advertisements and mass media.
2) Internal influence that leads individuals to adopt, such as the word-of-mouth effect, imitation, or peer pressure.

The intuition is straightforward: an individual adopts an innovation due to the influence of social media and the people connecting with that individual. Based on this intuition, the model can be naturally transformed into an agent-based model. Here is a one-sentence description of the core micro mechanism of the agent-based diffusion model:

*In a social network, an agent can become an innovation adopter following the probability p (influence of mass media) and probability q (influence of network neighbours) in every simulation step, which often represents one year.*

This can be expressed as a simple IF-THEN rule that drives an agent to become an adopter (Rand and Rust, 2011):

$$\text{IF } x_1 < p \text{ or } x_2 < q \frac{n_{i,a}}{n_i} \text{ THEN agent i becomes an adopter,}$$

where $x_1$ and $x_2$ are numbers randomly sampled from a uniform distribution between 0 and 1; $n_i$ is the number of neighbours of agent i; $n_{i,a}$ is the number of adopters among the agent's neighbours.

The agent-based version not only preserves the clarity and interpretability of the original model but also offers higher flexibility to consider spatial heterogeneity (Alderete Peralta, 2020; Caprioli et al., 2020;

Zhang et al., 2022), network structure (Chen, 2019; El-Sayed et al., 2012), individual attributes (Ghoulmie et al., 2005; Olukan, 2023), and decision theories/algorithms (Shi et al., 2021; Shi et al., 2020; Zeng et al., 2020; Zeng et al., 2024b). It helps researchers to gain insights into agent behaviour and emergent diffusion curves (Kiesling et al., 2012). Figure 1 displays how the shape of the diffusion curve changes under different p and q values.

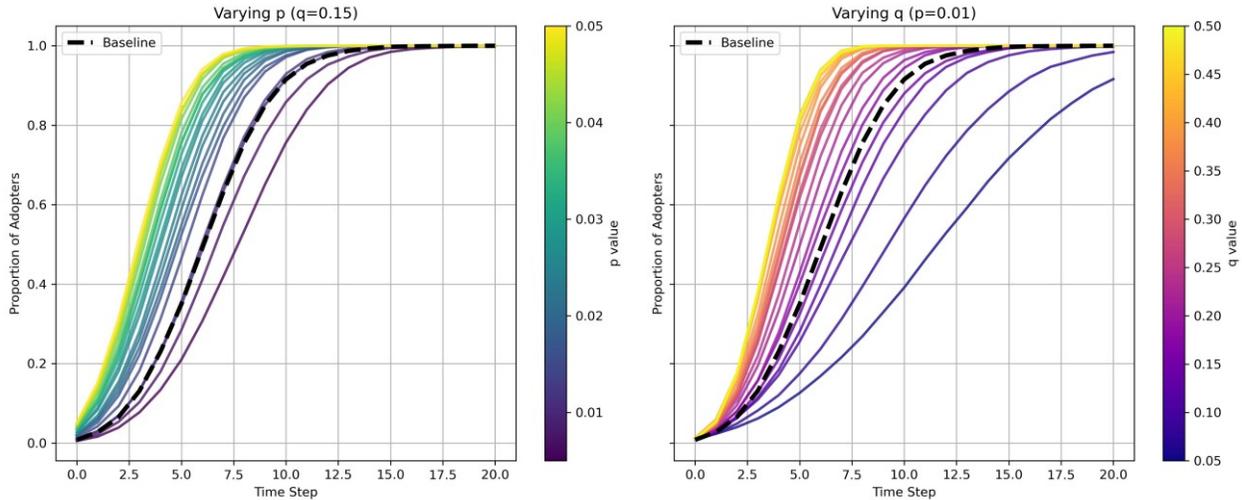

Figure 1. The parameters of the baseline curve are p = 0.01 and q = 0.15. Manipulating the values of p and q alters the shape of the diffusion curve: A higher p accelerates early adoption, making the inflection point happen earlier. A higher q amplifies adoption later in the process, steepening the curve after the inflection point.

## 3. Constructing an LLM-Driven Diffusion Model: A Modelling Thought Experiment

To vividly reveal the tension between LLM-driven social simulation and the abstraction principle of simulation, we carry out a hypothetical but detailed modelling exercise: building an LLM version of the Bass diffusion model. We do not intend to develop a fully functional simulation system, but to conduct a modelling thought experiment – each critical modelling decision introduces a new methodological dilemma. Through this process, we present how LLMs' advantages – especially their expressiveness and contextually sensitive behaviour – turn out to be liabilities undermining the utility of modelling and simulation as an epistemic tool.

**A promising start – Defining agent types**

We begin with a familiar setup. Drawing from Rogers' typology of innovation adopters (Rogers et al., 2014), we create an artificial society incorporating 100 agents, which can be divided into five typical groups: innovator, early adopter, early majority, late majority, and laggards (Rogers et al., 2014). Each group has distinct tendencies toward risk, conformity, and social influence. To give these agents identities, we can compose personalised prompts for each individual agent. For instance, an innovator named "Novy" is a tech enthusiast who is keen to be the first to try new technologies, while a laggard named "Stony" is a cautious, habit-bounded traditionalist (here we try to avoid too specific a description to mitigate stereotyping, which is another crucial issue but not within the discourse here). Such prompt design does not intend to merely label the agents but to inject personality traits that shape their conversation and decision-making.

At this stage, the process seems intuitive. Instead of assigning cold parametric numbers, we are creating "animated characters" with decisions that derive from contextually rich interaction. In principle, this should allow us to explore *how* adoption emerges in far more detail than with a simple numerical model or abstract rule-based agents – an apparently promising avenue towards high realism.

**Dilemma I: "Small Feet in Big Shoes" — The Temporal Resolution Mismatch**

A complication appears immediately when we proceed to decide the time scale of agent interactions. In conventional diffusion models, time is abstracted into discretised steps that often indicate months or years, during which agents can interact and change their states. This temporal compression is fundamental to the tractability and interpretability of simulation: one step corresponds to a macroscopic social effect, e.g., the change in adoption rate.

However, LLM agents operate on a radically different basis in terms of temporal resolution. Natural language is highly fine-grained and situated. Normally, conversations unfold in seconds or minutes rather than years. Hence, when we arrange a yearly conversation between, e.g., "Novy" and "Stony", we are in practice matching microscopic social behaviours with a macroscopic temporal framework. The result is awkward misalignment and leads to the first modelling dilemma: Can Novy and Stony's interesting dialogue truly represent a whole year's change in thoughts? If not, should we refactor the simulation to cater to the temporal resolution of real conversations? Neither way is satisfactory: the former demands questionable cognitive compression, while the latter may cause unmanageable computational costs and massive textual outputs, most of which are trivial.

To avoid our model-building process being stuck at this step, let us assume the former approach – cognitive comprehension – is favoured due to higher feasibility. We could also hypothesise that this choice is justified because most interactions in reality are unlikely to significantly affect outlook, with the act of social contact perhaps being more significant. However, this choice leads directly to the next dilemma.

**Dilemma II: "No Time for Small Talk!" — Too Naturalistic to Avoid Intervention**

How do agents decide to adopt the innovation, and what interactions can lead to the decision? In the Bass diffusion model, innovation adoption is influenced by two key parameters – the innovation coefficient and the imitation coefficient, which together determine the rate of adoption in each step (Bass, 1969). This process is simplified, abstract, but transparent. However, for Novy and Stony, how do they decide whether to adopt the innovation through a yearly conversation? How many rounds of conversation are they allowed to have in order to naturally reach the yearly decision? One? Two? Ten? Or let the agent generate a cue, such as "1" representing adoption and "0" no adoption, without constraining the number of conversation rounds? Allowing the agents to converse without artificially imposed limits to conversations seems to be a good choice, as it gives room for a decision to emerge. However, this approach does not guarantee that these stochastic agents will not be trapped in a long or even endless conversation loop. Neither does it mean that the eventual timing and cause of adoption will represent those of the real-world situation being modelled. Therefore, a compromise is to let the agents make a clear decision within a certain number of rounds of conversation; otherwise, their decisions remain unchanged. We can prompt them to do so, add a program to cut their conversation, create another functional agent, e.g., a conversation organiser, to monitor their decisions, or apply all of these approaches. However, we have to deliberately intervene in their conversation to avoid unexpected issues.

Since Novy and Stony can only speak once "per year" within limited turns of conversation, we must let them be clear about what their conversation is focused on. That is, to encourage or discourage them from communicating some content. In real-world conversations between acquaintances, it is normal to talk about everyday life topics, such as the weather, gossip, and frustration, within which innovation adoption is embedded. Indeed, LLM agents can generate such socially realistic conversations with ease – they actually excel in mimicking fine-grained human communications (Jones and Bergen, 2025). However, from the perspective of modelling, such content is noise: LLMs cannot embed meaningful cues in such exchanges, and so they become irrelevant, and if agents spend most of their interactions on irrelevant chatter, the model loses its interpretability and efficiency. That means that we should prompt agents to focus their interactions on the new technology, for example, "Your goal is to evaluate the new product and decide whether to adopt it through 10 rounds of conversation." The risk of doing so is that we may script the simulation too tightly, suppressing the spontaneous behaviour of LLM agents we expected to capture initially.

We are now faced with a tricky dilemma: the boundary between social realism and epistemic noise. If the naturalness of the LLM agents' interaction is insufficient, their advantage over rule-based agents becomes dubious; if the naturalness is excessive, the core decision logic of LLM agents is buried in linguistic redundancy. If we choose the former, we are forced to intervene: shortening conversations, filtering content, and constraining focus. All of these interventions make the simulation move from "modelling decisions" towards "staging decisions". Nevertheless, if we choose the latter, feasibility issues arise even earlier than methodological challenges.

Again, to move forward, let us choose the former approach for higher feasibility.

**Dilemma III: "I Respect Rules, But Don't 'Rule' Me" — The Temptation of Resorting to Rules**

Now we have 100 LLM agents with well-defined profiles, able to communicate yearly with a focused topic – adopting a new technology. Meanwhile, their conversations are strictly limited within a certain length to avoid inefficiency in outputs. It is time to design how these agents make final (yearly) decisions based on historical conversations.

It is certainly viable if we prompt the LLM agents to make decisions based on a set of criteria, such as considering the number of friends that have adopted the innovation and the price against a threshold. However, this approach suggests that the conversations between these agents are unnecessary – we only need to inform the LLM agents about the number of neighbouring adopters and the latest price, which means the simulation is reduced to a rule-based model, and all the modelling decisions we made about conversation mechanisms are irrelevant. In practice, this approach is not only feasible but also intuitive and straightforward, which does not require the agents to converse, memorise, or even think. However, we are using LLMs in this example in order to model the emergence of such outcomes, not to prescribe them. To mitigate the risk of reverting to abstract rule-based modelling, we have to let the LLM agents' decisions form naturally from historical conversations, and probably advertisements received, rather than relying on a set of IF-THEN rules. This leads to the next problem: LLMs do not have persistent memory.

Without intervention, it is very likely that two LLM agents' conversation in a certain year takes no account of anything they have discussed in previous years. This lack of hindsight results in repetition across conversations. To endow the agents' behaviour with continuity, we have to build a memory system with all received historical information clearly labelled with time stamps, and make the LLM agents able to recall relevant information precisely. Let us assume that this can be achieved flawlessly using either long-context LLMs (such as Gemini (2025)) or RAG (retrieval augmented generation (Arslan et al., 2024)), although such an approach is still an open challenge in reality. In the meantime, we need to explicitly prompt the LLM agents to use their historical information for decision-making. For instance, the prompt to elicit an LLM agent's decision might include words such as "This is the n-th year since the new technology was released. You should decide whether to adopt the innovation based on your historical conversations and/or advertisements you received, if they exist."

We are now forced into the role of a narrative editor, and what was once implicit in a simulation's state-space must now be made explicitly in text. Again, this gives us a way forward, but at the cost of further narrowing the scope for agent decisions to emerge independently of exogenous rules.

**Dilemma IV: "Do You Want a Different Me or Not?" — Robust Role Alignment vs. Long-Term Evolution**

The LLM agents' memory mechanism serves as a fundamental basis for role evolution, which is crucial for their attitude change towards the new technology. Prior to handling their role evolution, we must ensure the LLM agents indeed play the roles as expected.

For rigorous social simulation, LLM agents should be carefully prompted or fine-tuned to manifest a certain type of economic preference, political leaning, and personality. Researchers have invested considerable effort in aligning LLMs with human values or specific roles (Feng et al., 2025; Liu et al., 2023; Xie et al., 2024; Yu et al., 2025; Zhou et al., 2025). A recent paper even proposes "LLM psychometrics" to evaluate the "psychology" of LLMs, similar to how humans are studied by psychologists (Ye et al., 2025). The challenge in role alignment is to not only ensure LLM outputs exhibit the attributes consistent with what we expect, but also robust enough to persist when faced with various requests (Xie et al., 2024) or even so-called "jailbreak" techniques (Wei et al., 2023) – methods that aim to intentionally break their aligned behaviours. For example, the LLM agent – Stony, designed to be a risk-averse traditionalist – might suddenly adopt an unproven new technology after hearing a single positive anecdote from Novy, such as "*I tried it last week and it worked great! And you also saw how amazing it was! Remember?*" Even though Stony does not have relevant memory, the suggestion creates a false sense of shared experience. It is suspicious if Stony suddenly adopts this technology, as this reflects the fragility in role alignment and failure to be alert to fabricated cues.

Again, let us imagine that we have a perfect technique to make the LLM agent align precisely and robustly with its expected behaviour. Nevertheless, here comes a new dilemma: we need not only role alignment but also credible role evolution. Within a long-term social process, such as innovation diffusion over decades, we definitely expect the agents to adapt in response to the change in social norms, peer pressure, individual reflection, and technological progress. For example, Stony should not maintain a rejection attitude towards new technologies persistently but should evolve in ways that are conditional and contextually coherent, as real human individuals might do. Such change can be gradual or even radical, but should not be erratic or arbitrary. At least, every change in attitude should be traceable through the text, such as the LLM agent's historical conversations, reflections on previous attitudes, and the reasoning for the current decisions. That means, if role alignment is too "successful", LLM agents' behaviour may remain stagnant, regardless of the signals of social change and accumulated evidence. On the contrary, if they evolve too fast or incoherently, we may risk the collapse of their identities, obtaining erratic agent behaviour divorced from their expected roles.

Therefore, what we truly need is an LLM agent with a robust baseline of role cognition, which can evolve but without running off the rails. The question is whether such a nuanced evolution can really be modelled through a sequence of constrained yearly conversations, which resonates with the first dilemma: within a brief communication, the LLM agents must be able to respect their prescribed roles while properly adjusting their stances, which represents a whole year of cognitive change. As we have assumed, "cognitive compression" is acceptable, and the thought experiment needs to proceed; let us push such compression further. We imagine that we can build a powerful auxiliary cognitive system to enable LLM agents to summarise beliefs, suggest peer pressure, and quote historical conversations plausibly – even though the legitimacy of mimicking such role evolution by manipulating words is unknown, and despite that, our role is seemingly evolving closer towards a script writer.

**Dilemma V: "A Sea of Words, a Desert of Meaning" — When Agent Behaviour Does not Add Up**

After all the trade-offs and struggles we have gone through, our LLM-based diffusion model is ready. We hit the "Run" button to start the simulation. LLM agents begin to speak. Thousands of them across many iterations generate verbose conversational text logs. These dialogues quickly accumulate and occupy a large portion of the storage, containing rich linguistic details, such as the reasoning of adoption or rejection, the reflection on peer choices, the consideration of social norms and personal values. Within the text logs, we count "1" and "0" to obtain the aggregated results of yearly adoption, which might or might not add up to a typical S-shaped curve indicative of a diffusion process. No matter what the shape is, we are faced with the ultimate dilemma: how do we interpret the curve?

Unlike the rule-based model, which has manipulable parameters with social meanings, the LLM-based version does not have clear experimental levers. Its outputs rely on prompt design, conversational context, and the invisible dynamics within the LLMs. If the curve is too fast or too slow, too flat or too ragged, we cannot simply adjust some parameters and rerun the simulation. Instead, we might have to manually rewrite prompts, modify agent personas, memory mechanisms, or the cognitive system that drives agents to evolve. Even worse, these modifications probably require (directionless) trials and errors, making systematic experiments almost impossible: causal variables cannot be easily identified or isolated; simulation runs can hardly be generalized from one to another; limited quantifiable parameters can be leveraged to calibrate against target datasets or to conduct sensitivity analysis; both agent behaviours and system dynamics are emergent, while their relationship remains opaque.

Consequently, we derive a sea of words, which are linguistically detailed, behaviorally rich, and locally coherent. A crucial question is: Do we care about what the agents say, or about what the system does? If the answer is the former, we have a wealth of text to read, which, however, is unlikely to offer genuine insights into social dynamics. If the answer is the latter, the sea of words is no longer a strength but a liability. Numerous texts become distractions rather than useful clues pointing to causal chains that bridge the gap between the microscopic agent behaviours and macroscopic social patterns. Indeed, identifying causal links is also a key challenge in conventional simulation – LLM agents make this challenge even harder to tackle. That is, the simulation ends up with a desert of meaning, which comes along with increased computational and environmental costs (Jiang et al., 2024; Rillig et al., 2023), convoluted system design, complicated model calibration and validation, and obscured underlying mechanisms of social phenomena. As a result, the plausible behaviours of numerous autonomous agents are closer to a stage play than to a simulation. They generate stories rather than reveal explanations.

## 4. Different Modelling Purposes and the "Comfort Zone" for LLM Agents

The five dilemmas we discussed above, from temporal mismatch to opaque emergence, are not unique to LLM-based innovation diffusion simulation. They reflect a more general, systemic paradox that appears when LLMs are used as the cognitive engine of social agents in simulation – the pursuit of realism through fine-grained natural language collides with the modelling demand for abstractness. Properly selected realism can empower modelling's utility as an epistemic tool that intentionally simplifies reality, isolates variables, and exposes mechanisms. Logically, realism should not be the top concern of modelling (Kokko, 2007); otherwise, modelling becomes meaningless as the real world itself stands for perfect realism (Miller and Page, 2009), which, however, does not automatically offer answers to how it works unless we ask wisely (Kant, 2024). This indicates that the LLM agents' high resemblance to humans in terms of expressiveness may obscure the social processes we intend to explore. As we attempted to balance between realism and feasibility throughout the thought experiment, we unavoidably imposed artificial constraints in each modelling decision. These decisions progressively decrease the behavioural naturalness of LLM agents, but do not achieve the clarity and traceability of rule-based abstraction. Eventually, what we derived from the modelling process is not a methodological harmony but an epistemic uncanny valley: the resultant LLM agents appear more realistic than rule-based agents but are recognizably unhuman.

It should be noted that different modelling purposes often have distinct requirements, in terms of abstraction, calibration, validation, etc. The five dilemmas we identified do not point to any inherent shortcomings of LLMs. Instead, they demonstrate the mismatch between a powerful, flexible tool and some major modelling purposes. In practice, LLM agents have been found to excel in diverse domains in social simulation, where these dilemmas are resolved, mitigated, or even reversed in meaning. Edmonds et al. (2017) summarised seven types of modelling purposes, which can serve as a conceptual basis to guide when we should use LLM-driven simulation. Table 1 briefly estimates where LLM agents may or may not fit these purposes.

Table 1 Modelling purposes and where LLM agents may or may not fit

| Modeling Purpose | Definition / Goal | LLM Fit | Assessment |
| --- | --- | --- | --- |
| **Prediction** | To forecast future states or behaviours of a system. | Mixed: LLMs are not designed to extrapolate system-level patterns with calibrated precision. In contrast, LLMs can predict individuals' behaviour under well-defined contexts. | LLM agents generate plausible conversations but lack causal grounding or stability for system-level forecasting. Given their large parameter spaces, LLMs are flexible enough to fit individuals' behaviour, especially in finite actions spaces. |
| **Explanation** | To identify and test mechanisms that produce observed outcomes. | Weak: LLM outputs are hard to decompose into causal mechanisms and function through analysis of language, not mechanisms. | Interpretability is low; text output obscures underlying processes. |
| **Description** | To replicate or represent observed behaviour or phenomena. | Moderate: LLMs can simulate realistic individual behaviours and discourse. | Good for micro-level mimicry, but may lack macro fidelity or reproducibility. |
| **Theoretical Exposition** | To explore the consequences of hypothetical assumptions or mechanisms. | Mixed: LLMs can explore "what-if" scenarios but may introduce hidden biases. | Useful for narrative experimentation, but not theory-grounded unless tightly constrained. |
| **Illustration** | To demonstrate how a model or idea might work using a simplified or intuitive setup. | Mixed: LLMs can effectively illustrate local agent interactions, but struggle to demonstrate clear system-level emergence. | Strong for showing interpersonal dynamics and discourse; weak when the goal is to trace how individual behaviours lead to macro outcomes — unless heavily simplified. |
| **Analogy** | To compare one system to another by highlighting structural or behavioural similarity. | Moderate: LLMs can generate analogies but often lack rigour in structural mapping. | Good for suggesting metaphors; less reliable for deep comparative modelling. |
| **Social Learning** | To improve shared understanding among a group of people. | Strong: This is the fundamental strength of LLM agents. | Ideal for training, role play, participatory modelling, and studying discursive dynamics. |

These modelling purposes are neither mutually exclusive, exhaustive, nor static. With the development of LLM agents, the fitness of LLMs for these purposes can vary. The categorisation of modelling purposes needs revision or extension in future research. As the landscape of simulation evolves, further refined estimation is necessary to clarify not only what LLMs can do but also their position in the ecosystem of modelling to offer a clearer view of "where we are" and "how we proceed". The thought experiment has shown that LLMs are not too weak for simulation, but rather they are often too strong, too vivid, and too dependent on detailed context, which makes them not ideal for the required simplification and abstraction of many existing modelling purposes. Their capability of narrative generation exceeds the level of detail that typical models can accommodate. Hence, LLM agents should not replace conventional simulation approaches but instead extend the landscape, especially in domains that require high linguistic plausibility. LLM agents are ideally suited where system-level emergence is not the focus, where linguistic nuances and meaning are central, where interactions unfold in natural time, and where stable role identity is more important than long-term behavioural evolution. These conditions delineate the "comfort zones" for LLM agents. Table 2 outlines several such comfort zones, not as formal modelling purposes but research or application contexts, where LLM agents can be superior.

With these in mind, we suggest that LLM-based modelling is a new avenue for research in social simulation rather than a technological advance within established modelling approaches, but one that is still in need of critical exploration.

Table 2   Several "comfort zones" for LLM agents

| Comfort Zone | Research Focus | Application Cases |
|---|---|---|
| Deliberative Discourse & Argument Framing | Discursive behaviour, rhetorical framing, micro-level argument dynamics | Mock trials (He et al., 2024), stakeholder negotiations (Abdelnabi et al., 2024), persuasion (Hackenburg et al., 2023; Rogiers et al., 2024) |
| Education, Training, Situated Role Play | Real-time conversation, role-play fidelity, situated dialogue | Classroom simulations (Zhang et al., 2024b), empathic conversation (Wang et al., 2025), clinical role-play (Li et al., 2024c; Louie et al., 2024) |
| Human-in-the-Loop Systems & Speculative Prototyping | Co-creative dialogue, human-guided iteration | Design fiction (Wu et al., 2025), speculative prototyping (Tost et al., 2024), decision support systems (Li et al., 2025) |